\documentclass[aps,showpacs,preprint]{revtex4}
\usepackage{graphicx}

\begin{document}

\title{Magnetoresistance in semiconductor structures with hopping conductivity: effects of random potential
and generalization for the case of acceptor states}

\author{N.\,V. Agrinskaya}   \email{nina.agrins@mail.ioffe.ru}
\author{V.\,I. Kozub}
\author{A.\,V. Shumilin} \author{E. Sobko}
\affiliation{Ioffe Physical-Technical Institute of the Russian
Academy of Sciences, 194021, Saint Petersburg, Russia.}

\begin{abstract}
We reconsider the theory of magnetoresistance in hopping
semiconductors. First, we have shown that the random potential of
the background impurities affects significantly preexponential
factor of the tunneling amplitude which becomes to be a
short-range one in contrast to the long-range one for purely
Coulomb hopping centers. This factor to some extent suppresses the
negative interference magnetoresistance and can lead to its
decrease with temperature decrease which is in agreement with
earlier experimental observations. We have also extended the
theoretical models of positive spin magnetoresistance, in
particular, related to a presence of doubly occupied states
(corresponding to the upper Hubbard band) to the case of acceptor
states in 2D structures. We have shown that this mechanism can
dominate over classical wave-shrinkage magnetoresistance at low
temperatures. Our results are in semi-quantitative agreement with
experimental data.
\end{abstract}


\maketitle

\section{Introduction}
The problem of magnetoresistance in the hoping transport was
addressed decades ago. In particular, an interest to this topic
was related to important additional information provided by
corresponding experiments (including estimates of the localization
length). The most general and natural mechanism of positive
magnetoresistance of orbital nature is related to shrinkage of the
localized wave function by magnetic field; it was extensively
reviewed in \cite{Book}. Then another important mechanism of
orbital magnetoresistance was considered by Nguen, Shklovskii and
Spivak (for the review see \cite{SS}). It is related to a presence
of under-barrier scattering of hopping electrons by intermediate
hopping sites and to interference between different hopping
trajectories. Note that for the effective interference the
difference of lengths of different trajectories should not exceed
the localization length which restrict the location of the
trajectories to so-called "cigar region". A significance of this
mechanism was emphasized by the factor of exponentially-broad
scatter of hopping probabilities corresponding to different
"hopping resistors". As a result of "logarithmic averaging" over
different configurations the most important role is played by
those interference patterns where the total hopping probability
almost vanishes as a result of the destructive interference. The
magnetic field suppresses the interference and thus the average
effect is {\it negative}  magnetoresistance which appears to be
linear at weak magnetic field (although becoming quadratic at $H
\rightarrow 0$). An important features of the approach discussed
in \cite{SS} were as follows. First, the authors exploited an
assumption of a presence of many intermediate scatterers. Second,
following the theory \cite{LK}, the authors assumed the
preexponential factor to be equal to $\mu/r$ where $\mu$ is
scattering amplitude, $r$ is a distance between the hopping site
and the scattering center. The picture of interference
magnetoresistance considered in \cite{SS} was very rich including
a change of the sign of magnetoresistance, effects of spin glass
etc.

Somewhat later the problem was also discussed in
\cite{Shirmacher}, \cite{Raikh} where it was noticed that in
realistic situations the number of intermediate scatterers is
small and most probably equal to one or (in average) even less.
Another important ingredient of the paper \cite{Raikh} was a usage
of wave functions typical for Coulomb centers which have not
contained preexponential depending on $r$. In contrast to the
"scattering states" of \cite{SS} which contained preexponential
factors decaying with $r$, this situation can be specified as
"strong scattering case". Note that, although in \cite{Raikh} the
authors considered 2D hopping, they addressed to the case of
delta-doped layer and thus the asymptotic of the wave functions
was similar to 3D. The important result of theory suggested in
\cite{Raikh} was the following: the patterns of the negative
magnetoresistance were almost universal predicting the maximum
value of $\sim 0.6$ of the total resistance, and even the
combination of the negative magnetoresistanse and positive
wave-shrinkage magnetoresistance gave the maximum value of
negative peak  (with respect to average resistance) of around 40
percents.

Unfortunately, these predictions for "strong scattering case" were
not in a good agreement with experiment. First, in most of
experimental studies the effect of negative magnetoresistance have
not exceeded 10 percents and typically was around several
percents. Then, it was shown \cite{ours1} that in 3D
semiconductors the negative magnetoresistance is suppressed with a
decrease of temperature after the crossover from Mott-type hopping
(at higher temperatures) to Efros-Shklovskii hopping over the
states within the Coulomb gap. In the paper \cite{ours1} we
explained such a behavior as a result of a decrease of
concentration of the scattering centers within the Coulomb gap.
However our calculations were based on the assumption that the
preexponential factor of the wave functions asymptotic
corresponded to scattering states of \cite{SS} ("weak scattering
case") rather than to hydrogen-like asymptotics exploited in
\cite{Raikh}. Later \cite{ours2} we have also demonstrated that to
fit the experimental data one should also take into account spin
mechanisms of magnetoresistance. The first one, considered in
\cite{Spivak}, is based on the fact that the intermediate
scatterer should be occupied to produce a negative scattering
amplitude. Thus the interference depends on the mutual orientation
of the spin of the hopping electron and of the spin of scattering
center. Without external magnetic field only one half of the
configurations gives an interference. In magnetic field all
localized spins are aligned which increases the role of
interference and, correspondingly, leads to an increase of
resistance.

Another spin mechanism of positive magnetoresistance was first
considered in Ref. \cite{Kamimura} and then was studied in detail
in \cite{Matveev}. It is related to a presence of doubly-occupied
hopping sites (corresponding to the upper Hubbard band). Due to
spin correlations on these sites requiring s-pairing of the spins
(recall that we consider here electron rather than hole hopping)
some hopping transitions are suppressed in magnetic field (like
ones from single-occupied site to single-occupied site).

As it was mentioned above, the incorporation of all of the
relevant factors
 allowed us to reach a quantitative agreement between the
theoretical model and experimental data. However basing the
scattering state asymptotic we exploited an assumption of
correlated impurity configurations which had no solid theoretical
prove.

Another important request to the theory of hopping
magnetoresistance was related to 2D hopping. As we have mentioned
above, the theoretical model of \cite{Raikh} exploited 3D
localized wave functions which do not hold for typical experiments
for doped quantum wells where the wave functions have 2D
character. Then, we should mention a new important experimental
results \cite{ours3},\cite{ours4}   obtained for selectively-doped
quantum well structures where both centers of the wells and
centers of the barriers were doped ensuring a formation of the
upper Hubbard band. These structures demonstrated a suppression of
negative magnetoresistance with a decrease of temperature for the
samples with higher degree of doping. Although we attempted to
explain this behavior in a similar way as for 3D structures in
\cite{ours1}, it hardly works because of an important difference
between 2D and 3D physics.

In what follows we will give a consistent description of
magnetoresistance in both 3D and 2D structures including different
orbital and spin mechanisms. An important conclusion of ours is
that in most occasions one deals with a "weak scattering case"
rather than with "strong scattering case". If we are restricted to
the lower Hubbard band, the decisive factor is related to the
presence of charged centers outside of the "cigar region" not
involved into interference. The random potential imposed by these
centers restricts the extension of the hydrogen-like asymptotics
of the scattering centers up to the distance to the closest
charged center while outside this region the preexponential of the
asymptotics appears to be similar to the one for the potential
well case ("weak scattering limit"). For the case of the states
within the upper Hubbard band an additional factor is related to
the non-Coulombic potential of the scattering center which is also
of a short-range character. The resulting picture of hopping
magnetoresistance appears to be different from the one suggested
in \cite{Raikh} (based on the pure Coulomb wave functions) and
from the one of \cite{SS} (exploiting the assumption of large
number of intermediate scatterers). We also emphasize a role of
spin mechanisms of positive magnetoresistance which can dominate
over wave-shrinkage magnetoresistance at low temperature. In this
concern a special analysis is given to spin mechanisms for
acceptor centers which have an important differences with respect
to the earlier discussed case of donor impurities.

\section{Negative magnetoresistance in 3D case.}
Let us consider negative magnetoresistance in 3D case. As it was
mentioned above, an important ingredient to be included with
respect to the previous studies is a random potential imposed by
the intermediate charged centers (including both donors and
acceptors).

We shall start from a solution of a Schrodinger equation
\begin{equation}
\label{Shred3D} - \frac{\hbar^2}{2m}\Delta \Psi +U_0({\bf r})\Psi +
U({\bf r})\Psi = E \Psi.
\end{equation}
Here $U_0({\bf r})$ is the potential of impurity ($U_0 = -\alpha/r$
in the case of hydrogen-like impurity level) and $U({\bf r})$ is
random potential that comes from the charged centers mentioned
above, $m$ is the electron mass in the conduction band (or a hole
mass in valence band) and $E$ is exact electron energy (we consider
$|E| \gg U({\bf r})$).

Because of the fact that typical hopping lengths are much larger
then characteristic localization length $a$, we can solve
(\ref{Shred3D}) at $r \gg a$. Moreover, the typical hopping length
$r_h$ appears to be much larger than typical distance between
charged centers which can be roughly estimated as $n^{-1/3}$ where
$n$ is a dopant concentration. Indeed, for 3D variable range
hopping of the Mott type
\begin{equation}
r_h = \xi a, \hskip1cm \xi = \left(\frac{T_0}{T}\right)^{1/4},
\end{equation}
where
\begin{equation}
T_0 \simeq \frac{21}{ga^3} ,
\end{equation}
where $g \sim n/{\cal E}_B$ is the density of states, ${\cal E}_B$
being Bohr energy. Thus one obtains
\begin{equation}
r_hn^{1/3} \sim \left(\frac{21 a n^{1/3}{\cal E}_B}{T}\right)^{1/4}
\end{equation}
that is even for the Mott law $r_hn^{1/3} >> 1$. The more so it
holds for the Coulomb gap regime where $r_h$ strongly exceeds the
corresponding values for the Mott regime.

Thus we are interested in asymptotics of the wave functions at
distances much larger than $n^{1/3}$ which for moderately
compensated material can be considered as the correlation length of
the random potential imposed by the charged centers. If so, we can
make an important conclusion. Namely, the random potential $U$ is
formed by the long-range Coulomb centers and in this sense the
potential produced by the "parent" (for the considered wave
function) impurity at distances larger than $n^{-1/3}$ makes no
difference with respect to potential produced by other charged
centers. In other words, for $r > n^{-1/3}$ one should not
discriminate between $U$ and $U_0$ and should assume that the
resulting potential $U$ has a spatial average equal to zero.

Having in mind that $r >> a$  we approach the problem of the
asymptotics of the wave function by means of the WKB method.  We
introduce the function $\phi$ with an account of the normalization
factor for the function $\Psi$ as: $\Psi = (\pi
a^3)^{-1/2}\exp(-\phi/\hbar)$. The Shroedinger equation in 3D case
leads to
\begin{equation}
\label{3DVKB1} \phi'^2 - \hbar(\phi'' + 2\phi'/r) = 2m(U({\bf r}) +
|E|).
\end{equation}
We will expand this equation into series with respect to $\hbar
\rightarrow 0$. In zero order we have
\begin{equation}
\label{3DVKB0O} \phi_0' = \sqrt{2m(U+|E|)}
\end{equation}
This order gives us the exponent. To get the pre-exponent factor
we should use the first order of perturbation theory. Here we have
in mind that the function $\phi_0'$ at large $r$ is actually a
constant - the more so that its linear expansion in $U(r)$ is
averaged out.

\begin{equation}
\label{3DVKB1O} \phi_1' =  \frac{\hbar}{r}
\end{equation}
Accordingly the expression for $\phi$ up to the first order is
\begin{equation}
\label{3DVKB1Phi} \phi = \int \sqrt{2m(U+|E|)} dr +  \hbar\ln
r/{r_{min}}.
\end{equation}
where $r_{min} \sim n^{-1/3}$. And finally the wavefunction $\Psi$
is
\begin{equation}
\Psi = \exp\left( - \frac{1}{\hbar}\int \sqrt{2m(U+|E|)} dr
\right)\frac{r_{min}}{r(\pi a^3)^{1/2}}
\end{equation}

Having in mind the considerations given above, we can average
$$
(1/\hbar)\int \sqrt{2m(U+|E|)} dr = kr
$$
where
$$
k=(1/\hbar)\left< \sqrt{2m(U+|E|)}  \right>.
$$
 Note that $k$ differs from $k_0 =
\sqrt{2m|E|}/\hbar$ only in second order of $U/|E|$ ($\propto
U^2/E^2$), as the mean value $\left< U \right>$ is zero. Also we
neglect $U$ in the pre-exponent factor.

So at distances from the scattering center larger than the
correlation length of the random potential (assumed to be equal to
average distance between the charged centers) the wave function
asymptotics has a preexponential factor $\propto r^{-1}$ which
agrees with the scheme exploited in \cite{SS}, \cite{ours1},
\cite{ours2} for 3D hopping.

Now, following approaches \cite{SS}, \cite{ours1}  let us estimate
the hopping probability between the sites 1 and 2 in a presence of
intermediate "scattering center" with an account that the energies
of the centers obey a relation $|E_3| \gg |E_1|,|E_2|$ as
\begin{equation}
\label{2DMRP1} P \propto |J_1+J_2|^2, \quad J_1 = I_{12}, \quad
J_2 = -\frac{I_{13}I_{32}}{|E_3|}.
\end{equation}
Here $J_1$ and $J_2$ are hoping amplitudes related to direct and
scattered path correspondingly. Note that the destructive
interference (leading to negative magnetoresistance) implies that
$E_3 < 0$) which means that in the equilibrium the scattering site
is occupied.

The energy overlapping integrals are given as
\begin{equation}
I_{ij} = {\cal E}_B\frac{r_{min}}{r_{ij}}\exp(-r_{ij}/a)
\end{equation}
where we have assumed that $r_{ij} > r_{min} = n^{-1/3}$; ${\cal
E}_B$ being the Bohr energy. Without a magnetic field this
amplitudes are real. Though in the magnetic field their phases are
different and hoping probability is
\begin{equation}\label{varphi}
\label{2DMRP2} P \propto |J_1 + J_2 e^{i\varphi}|^2.
\end{equation}
Here phase difference $\varphi$ is equal to $\varphi = 2 \pi
\Phi/\Phi_0$, where $\Phi$ is the magnetic flux through the
surface bounded by hoping paths. $\Phi_0$ is the elementary
magnetic flux. Accordingly, the interference magnetoresistance for
the situation $\varphi < 1$  can be given as
\begin{equation}
\label{2DMR1} \ln \frac{r(H)}{r(0)} \propto - \left< \int d E_3
g(E_3) \int \ln \left[1+J_1(J_1-J)\frac{\varphi^2}{J^2}\right] d^3
r_3 \right>,
\end{equation}
Here $J = J_1 + J_2$, $g$ is the density of states and $r_3$ is
scatterer position. Angle brackets corresponds to the ensemble
average. We consider magnetoresistance to be determined over hops
with small $J$, so we neglect the term $J_1/J$ in (\ref{2DMR1}) and
get
\begin{equation}
\label{2DMR2} \ln \frac{r(H)}{r(0)} \propto - \left< \int d E_3
g(E_3) \int \ln \left[1+\frac{J_1^2\varphi^2}{J^2(r_3, E_3)}\right]
d^3 r_3 \right>.
\end{equation}
To obey $J_1 \simeq J_2$ one, first, should have $r_{12} \simeq
r_{13} + r_{23}$ with an accuracy of the order of localization
length $a$. Then, having in mind the preexponential factors one
notes that for small $r_{min}$ the only possibility to obey the
relation is to have one of the distances, $r_{13}$ or $r_{23}$ to
be small. We will assume that it holds for $r_{23}$ which is
estimated as
\begin{equation}\label{r23}
r_{23} \sim r_{min}\frac{{\cal E}_B}{E_3}
\end{equation}
Let us chose the surface at which $J = 0$ and transform an
integration over $r_3$ in a way $d^3 r_3 \rightarrow d^2{\cal R} d
R_{\perp}$ where ${\cal R} $ is the coordinate on the surface in
question while $R_{\perp}$ is a coordinate along the normal to the
surface where we assume that $R_{\perp} = 0$ corresponds to $J=0$.
In the lowest order in $R_{\perp}$ we have $J =
(dJ/dR_{\perp})R_{\perp}$. As it is seen, the integration of the
logarithm term over $R_{\perp}$ gives
$$ \frac{\varphi J_1}{d J/d R_1}$$ Finally, the integration of the
factor $$ \frac{J_1}{dJ/dR_{\perp}}$$ over $d^2 {\cal R}$
approximately gives a volume accessible for the site 3.
 Note that we have $r_{13} + r_{23} \leq r_{12} + a$,
thus the projection of $\bf r_{23}$ to the plane normal to $\bf
r_{12}$ should be less than $(r_{23}a)^{1/2}$. As a result, the
integration over the spatial coordinate $\bf r_3$ gives
\begin{equation}
\sim \frac{r_{min}^2 {\cal E}_B^2}{E_3^2}a\varphi
\end{equation}
In its turn, the area of the interference loop (entering the
estimate of $\varphi$) is
\begin{equation}
r_{12}\left( r_{min}\frac{{\cal E}_B}{E_3}a\right)^{1/2}
\end{equation}
Note that these estimates actually hold for all accessible values of
$r_{23}$ up to $r_{23} \sim r_{12}/2$. The final result depends on
the behavior of $g(\varepsilon)$. For $g = const$ (Mott-type
hopping) the integration over $E_3$ is naturally controlled by the
lower limit which accordingly to Eq. \ref{r23} corresponds to the
larger possible value of $r_{23} \sim r_{12}/2$. In this case the
r.h.s. of Eq.\ref{2DMR1} is $\propto r_h^{5/2}$ where $r_h \sim
r_{12}$.

In contrast, for the Coulomb gap hopping the integration over
$E_3$ is controlled by the upper level, $E_C$, corresponding to
the edge of the Coulomb gap. In this case r.h.s. of Eq.\ref{2DMR1}
is $\propto r_h$ since the value of $r_{23}$ does not depend on
$r_h$.

Now let us consider a combination of NMR with a positive
magnetoresistance related to wave function shrinkage which can be
estimated as
\begin{equation}\label{shrinkage}
\ln \frac{\rho(H)}{\rho (0)} = \left(\frac{H}{B}\right)^2
\end{equation}
where
\begin{equation}
B^2 = \frac{\alpha c^2 \hbar^2}{r_h^3 a e^2}
\end{equation}
Here $\alpha$ is a numerical parameter resulting from he
percolation theory; for Mott type hopping $\alpha \sim 400$
\cite{Book}  while for the Coulomb gap hopping different sources
give $\alpha \sim 300$ and $\alpha \sim 700$.

In its turn, NMR can be rewritten as
\begin{equation}
\ln \frac{\rho(H)}{\rho (0)} = k \frac{H}{B}
\end{equation}
where
\begin{eqnarray}
k = g_M {\cal E}_B r_{min}r_h 2a\alpha^{1/2} \hskip1cm {\rm  Mott
\hskip0.5cm law}
\nonumber\\
k = \frac{\kappa^3}{e^6}\frac{r_{\Delta}^{5/2}}{r_h^{1/2}} a E_C^3
2{\alpha}^{1/2} \hskip1cm {\rm  ES \hskip0.5cm law}.
\end{eqnarray}
Here $\kappa$ is the dielectric constant, $r_{\Delta}$ is the
typical hopping length for the states corresponding to the edge of
the Coulomb gap while $E_C$ is a width of the Coulomb gap.
One sees that as a result we have minimum of resistance,
\begin{equation}\label{min}
H_{min} = \frac{k}{2}B, \hskip1 cm \ln \frac{\rho(H_{min})}{\rho
(0)} = - \frac{k^2}{4}
\end{equation}
It is seen that the value of $H_{min}$ decreases with a
temperature decrease irrespectively to the type of the variable
range hopping. At the same time for samples corresponding to Mott
law the absolute value of resistance in minimum {\it increases}
with a temperature decrease while for the case of the Coulomb gap
hopping it decreases with temperature decrease.

\section{Negative Magnetoresistance in 2D.}

Let us now approach the problem of negative magnetoresistance in
the 2D structure where impurity wave functions are quantized in
the orthogonal to impurity plane direction. First we will consider
the case when we deal only with single occupied or empty impurity
centers (as it was done above for 3D case).

Let us start with approximation of 2D impurity wave function in the
$r \gg a$ limit. Analogously to previous case we neglect $U_0({\bf
r})$ and introduce function $\phi$ as $\Psi = (\pi a^2)^{-1/2}
\exp(-\phi/\hbar)$. The corresponding WKB equation is
\begin{equation}
\label{2DVKB} \phi'^2 - \hbar(\phi'' + \phi'/r) = 2m(U({\bf r}) +
|E|).
\end{equation}

Following the same procedure as was applied for 3D case we obtain
\begin{equation}\label{Psi2}
\label{2DVKBf} \Psi = \exp\left( - \frac{1}{\hbar}\int
\sqrt{2m(U+|E|)} dr \right)\left(\frac{r_{min}}{r\pi
a^2}\right)^{1/2}.
\end{equation}
Analogously to 3D case this wave function is nearly equal to the
potential well wave function $ \propto \exp(-kr)/\sqrt{r}$ where
$k=\left<\sqrt{2m|E| + U} \right>/\hbar$ which differs from $k_0 =
\sqrt{2m|E|}/\hbar$ only in the second order of $U/E$.

Now let us consider negative magnetoresistance related to the
interference contribution to the hopping probability. Following
the same lines as for 3D case we obtain the equation similar to
\ref{2DMR2} except that the integration is over $d^2r_3$ and the
density of states $g$ also corresponds to 2D. An important
difference is related to the fact that now the value of $J$
vanishes at
\begin{equation}
r_{23} = r_{min}({\cal E}_B/E_3)^2
\end{equation}

With a similar transformation of the variables the integration of
the logarithmic term over the coordinates gives
\begin{equation}
\left(\frac{r_{min}{\cal E}_B^2}{E_3^2}\right)^{3/2}a^{1/2}
\varphi
\end{equation}
while the effective loop area is
\begin{equation}
r_{12}\left( r_{min} \left(\frac{{\cal E}_B}{E_3}\right)^2
a\right)^{1/2}
\end{equation}
Since in 2D in the Coulomb gap regime $g \propto \varepsilon$ one
notes that irrespectively to the hopping law the integration over
$E_3$ is controlled by the lower possible values of $E_3$ leading
finally to the estimates of $r_{23} \sim r_{12}$. Thus one obtains
\begin{eqnarray}\label{k2}
k_2 = g_M {\cal E}_B r_{min}^{1/2}r_h 2a^{1/2}\alpha^{1/2} \hskip1cm
{\rm
Mott \hskip0.5cm law} \nonumber\\
k_2 = \frac{\kappa^2}{e^4} {\cal E}_B^2 r_{min}r_h^{1/2}
2a^{1/2}\alpha^{1/2} \hskip1cm {\rm ES \hskip0.5cm law}
\end{eqnarray}
Thus, at is seen, for the situation considered above in 2D the
only combination of interference NMR and wave-shrinkage PMR can
not lead to a suppression of NMR with a decrease of temperature
(at least for low temperature limit of linear NMR) since for both
laws the temperature derivative of $k$ stays to be negative.

An important feature of the 2D quantum well structure is an easy
possibility to have an occupation of the upper Hubbard band.
Namely, if we dope not only the well regions, but also the barrier
regions, the carriers from the barriers are captured by the wells
and can form doubly occupied states. It was this situation which
was realized in our experiments described in \cite{ours3},
\cite{ours4}. Since in these experiments we dealt with GaAs/AlGaAs
structures of quantum wells with p-doping by Be, here we will also
imply acceptor centers.

In our experiments the central  regions of both wells and barriers
were nearly equally doped by acceptor impurity Be. Thus the holes
from the barriers have a possibility to occupy the second position
for the acceptor in the wells forming the upper Hubbard band.
However for the hole there was another possibility - to stay
around its native acceptor in the barrier forming single-occupied
center which we will denote as  $\tilde A^0$. The corresponding
scenario was first discussed in \cite{Larsen}. As a result, at the
Fermi level we have centers with different occupation numbers - at
least, $A^+$ (doubly occupied). $A^0$ (single occupied)  $\tilde
A^0$ (holes bound to the barrier acceptor)  and $A^-$ (empty
barrier acceptor with no hole around).

The possibility for the hole to form $A^+$ or $\tilde A^0$ center
depends on relation between the binding energies of these centers,
$U_b$ and $\tilde U_b$. In particular, if $U_b > {\tilde U_b}$,
then all the barrier acceptors form $A^-$ centers while all the
acceptors in the well form $A^+$ centers. However for our
experiments of the distance between the barrier acceptor and the
interface between the barrier and well was not large and we expect
${\tilde U_b} > U_b$. In this case the probability to form $A^+$
center depends on the distance between the barrier acceptor and
the closest acceptor in the well. Indeed, the formation of $A^+$
center profit from the interaction between $A^+$ center and $A^0$
center \cite{Larsen}.

Here we assume that some holes from the barrier are still coupled
to their parent acceptors ($\widetilde{A}^0$ centers) and some are
localized on the acceptors in the well ($A^+$ centers). According
to charge conservation the number of $\widetilde{A}^-$ centers
(that are free $\widetilde{A}^0$ centers) is equal to the number
of $A^+$ centers.
\begin{equation}
\label{NA0A+} N(A^+) = N(\widetilde{A}^-).
\end{equation}

In addition, we believe that there exists a random potential that
overlap the energies of different types of centers. If the
variances of $\widetilde{A}^0$ and $\widetilde{A}^-$ energies are
equal, (\ref{NA0A+}) leads to equal densities of states for
$\widetilde{A}^0$ and $\widetilde{A}^-$ at the Fermi level. For
our purpose we assume that this densities of states are at least
comparable.

As for the negative magnetoresistance for the upper Hubbard band,
it can be considered in the same way as for the lower Hubbard band
discussed above. Note that the scattering potential strongly
decays with distance $U_0 \propto r^{-4}$ and thus the
corresponding asymptotics of the wave functions are similar to the
one given by Eq.\ref{Psi2} but one should take $r_{min} = a$.

\section{Spin mechanisms of magnetoresistance for acceptor states}

We shall start from the mechanism of spin magnetoresistance first
suggested in \cite{Spivak} which seems to be especially important
for acceptor dopants. It is related to the fact that interference
can occur only if spin states of the final states for both
tunneling paths coincide. For 3-cite configuration we discuss it
means that the initial and intermediate centers should have the
same spin projections (we remind that for destructive interference
in question the energy of intermediate center should be negative,
i.e. at the equilibrium this center should be occupied). For the
case of acceptor states corresponding to the lower Hubbard band
the corresponding configuration is in our case ${\tilde A}^0 -
{\tilde A}^0 - A^-$ where the role of intermediate center is
played by the site ${\tilde A}^0$. Since the hole has spin 3/2, we
have 4 projections of the spin and thus the probability  for two
sites ${\tilde A}^0$ to have the same spin projections is $P(H=0)
= 1/4$.  However in strong magnetic field the site spins are
aligned and spin does not affect the (destructive) interference,
that is in this case $P(H \rightarrow \infty) = 1$. Thus an
increase of the magnetic field leads to an enhancement of
destructive interference which means positive magnetoresistance
which was noted in \cite{Spivak}.

In a presence of the states representing the upper Hubbard band
(in our case of $A^+$ and $A^0$ states) the situation is somewhat
more complicated. In particular, it is related to the fact that
the spin structure of doubly occupied $A^+$ center is more complex
than for single occupied site. In particular, the total spin of
$A^+$ center is 2 (see \cite{A+}) and the possible spin states of
$A^+$ center are the following:
$$
\left|J=2, J_z = -2 \right> =
\frac{1}{2}\Psi^{(1)}_{-1/2}\Psi^{(2)}_{-3/2} -
\frac{1}{2}\Psi^{(1)}_{-3/2}\Psi^{(2)}_{-1/2}
$$
$$
 \left|J=2, J_z = -1\right> = \frac{1}{2}\Psi^{(1)}_{1/2}\Psi^{(2)}_{-3/2} -
\frac{1}{2}\Psi^{(1)}_{-3/2}\Psi^{(2)}_{1/2}
$$
$$
 \left|J=2, J_z=0\right> = \frac{1}{4}\Psi^{(1)}_{3/2}\Psi^{(2)}_{-3/2}
 + \frac{1}{4}\Psi^{(1)}_{1/2}\Psi^{(2)}_{-1/2} - \frac{1}{4}\Psi^{(1)}_{-3/2}\Psi^{(2)}_{3/2}
 - \frac{1}{4}\Psi^{(1)}_{-1/2}\Psi^{(2)}_{1/2}
$$
$$
 \left|J=2, J_z = 1\right> = \frac{1}{2}\Psi^{(1)}_{3/2}\Psi^{(2)}_{-1/2} -
\frac{1}{2}\Psi^{(1)}_{-1/2}\Psi^{(2)}_{3/2}
$$
$$
 \left|J=2, J_z = 2\right> = \frac{1}{2}\Psi^{(1)}_{3/2}\Psi^{(2)}_{1/2} -
\frac{1}{2}\Psi^{(1)}_{1/2}\Psi^{(2)}_{3/2}
$$
where $\Psi^{(1,2)}_s$ - are the wave functions characterized by
given spin projections of the two holes. Basing on these
considerations one can show that for the destructive interference
involving purely the states of the upper Hubbard band, that is for
configurations $A^+ - A^+ - A^0$, $P(H=0) \sim 1/4$ while $P(H
\rightarrow \infty) = 1$

If the states of both of the Hubbard bands coexist at the Fermi
level, it can be estimated that the average statistical factor
$P(H=0)$ is still of the order of 1/4, although its value at
strong fields, $P(H \rightarrow \infty)$ appears to be somewhat
smaller than unity.

At weak magnetic fields one expects a degree of spin alignment  to
be $\propto (\mu g H)^2/T^2$ and thus the statistical factor is
equal
\begin{equation}
P(H) \simeq P(H = 0) + \alpha \left( \frac{\mu gH}{T} \right)^2
\end{equation}
Here the coefficient $\alpha$ according to more detailed
statistical calculations which we are going to present elsewhere
can be estimated to be of the order of 1/2.

Since $P$ describes probability of the destructive interference,
one concludes that at weak fields the positive magnetoresistance
resulting from statistical factor $P$ is quadratic in terms of
magnetic field. It can be estimated as follows:
\begin{equation}\label{statquadrat}
\ln \frac{R(H)}{R(0)} = \alpha \left( \frac{\mu gH}{T} \right)^2
 \left| \frac{\Delta R_{sat}}{R(0)}\right|
\end{equation}
where and $\Delta R_{sat}$ is the saturation value for the
interference contribution to resistance with no account of spin
degrees of freedom which is achieved when the phase $\varphi$ in
Eq.\ref{varphi} exceeds $2\pi$. As it can be estimated, the ratio
$\Delta R_{sat}/R(0)$ is $\propto r_h$ for Mott hopping and
$\propto r_h^{1/2}$ for the Coulomb gap hopping.

In its turn, it coexists with linear negative magnetoresistance
(of orbital nature) which at relatively weak fields can be
estimated as
$$ P(H = 0) \Delta R_{sat}\frac{H}{H_{sat}} $$

As it is known, if at the Fermi level the states of the lower and
the upper Hubbard bands coexist, there also exists a specific spin
mechanism of positive magnetoresistance first considered in
\cite{Kamimura} (and later discussed in detail in \cite{Matveev})
for n-type 3D structures. In such structures one deals with $D^0$
(occupied donors), $D^-$ (doubly occupied donors) and $D^+$ (empty
donors). Without external magnetic field the following
configurations of hops are possible: $D^- \rightarrow D^0$, $D^0
\rightarrow D^0$, $D^0 \rightarrow D^+$, $D^- \rightarrow D^+$. In
the magnetic field the spins of $D^0$ centers are polarized and
thus the hops $D^0 \rightarrow D^0$ are forbidden (since in the
final state of the second site corresponding to $D^-$ the spins
should be in opposite directions). In the same way the transitions
$D^- \rightarrow D^+$ are also suppressed. Thus the resistance
increases as a result of application of external magnetic field.

In our case of p-type structures the situation appears, again,
more complicated due to more complex structure of $A^+$ centers.
However, in general, the considerations given in \cite{Kamimura},
\cite{Matveev} still hold.  Basing on the calculations similar to
given in \cite{Matveev} one obtains for weak field limit $\mu g H
< T$ the following estimate:
\begin{equation}\label{spin}
\ln \frac{R(H)}{R(0)} = C F \left( \frac{g \mu_b H}{T}\right)^2
\end{equation}
where $C \sim 1/3$,
\begin{equation}
F = \frac{2g_{l}g_{u}}{( g_{l}+g_{u})^2}
\end{equation}
while $g_{l}$, $g_{u}$ are the densities of states of the lower
and upper Hubbard bands. Note that for the low concentration of
dopants $g_u$ is controlled by the concentration of $A^+$ centers
while $g_l$ - by the concentration of $A^-$ centers and thus $g_l
= g_u$.  At stronger magnetic fields when $\mu g H > T$, the
corresponding contribution to magnetoresistance still increases
with magnetic field increase until $\mu g H$ reaches the value
$\xi T$ and then saturates \cite{Kamimura}, \cite{Matveev}.

One notes that at low enough temperatures the positive
magnetoresistance of the spin nature suggested in \cite{Kamimura}
can exceed the wave shrinkage magnetoresistance. At the same time
this contribution at relatively weak fields when $\mu g H < T$  is
expected to be comparable to the spin magnetoresistance resulting
from interference term discussed above. Summarizing the both spin
contributions to quadratic magnetoresistance we estimate the
coefficient $k$ resulting from the similar parametrization of the
positive quadratic and linear negative magnetoresistance as was
done above:
\begin{eqnarray}\label{kkamimur}
k_2 = g_M E_B r_{min}^{1/2}r_h 2a^{1/2}\beta \hskip1cm {\rm
Mott \hskip0.5cm law} \nonumber\\
k_2 = \frac{\kappa^2}{e^4}E_B^2 r_{min}r_h^{1/2} 2a^{1/2}\beta
\hskip1cm {\rm ES \hskip0.5cm law}\nonumber\\
\beta = P(H = 0)
\frac{T}{g\mu_B}\frac{r_h^{3/2}a^{1/2}e}{c\hbar}(CF + \alpha
\frac{\Delta R_{sat}}{R(0)} )^{-1/2}
\end{eqnarray}
Thus, as it is seen, for the Mott case at $T \rightarrow 0$ $k
\propto T^{1/3}$ while for the ES case it is $\propto T^{1/4}$.

Note that in our calculations we assumed that the value of
$H_{min}$ still corresponds to linear behavior of negative
magnetoresistance which means that the magnetic flux through the
interference area is much less than magnetic flux quantum
$\Phi_0$. The critical field $H_{sat}$ corresponding to a
crossover from the linear behavior to saturation of the negative
magnetoresistance is given as
\begin{equation}
H_{sat} \simeq \frac{\Phi_0}{2 \pi r_h^{3/2}a^{1/2}}
\end{equation}

Correspondingly, if $H_{min}$ given by Eq.\ref{min} appears to be
larger than $H_{sat}$ our calculations given above are invalid and
one should compare positive magnetoresistance with saturated
negative magnetoresistance rather than with linear negative
magnetoresistance. One notes that in contrast to linear
magnetoresistance which is proportional to the area of the
interference loop for the saturation magnetoresistance this
proportionality is omitted. As a result, as it was noted above, the
temperature dependence of the saturation value of negative
magnetoresistance $\Delta R_{sat}/R(0)$ results from factors
$\propto r_h$ for Mott hopping and $\propto r_h^{1/2}$ for the
Coulomb gap hopping. It is seen that the corresponding increase of
the saturation magnetoresistance with temperature decrease is much
weaker than increase of the positive magnetoresistance. Then, in the
case $H_{min} > H_{sat}$ it is the value of $H_{sat}$ which
corresponds to minimal resistance since it separates a region of
resistance decrease due to negative magnetoresistance and resistance
increase due to positive magnetoresistance. However at this
situation spin magnetoresistance (\ref{statquadrat}) is also
saturated so the temperature behavior of positive magnetoresistance
is related to (\ref{shrinkage}) and (or) to (\ref{spin}).

In its turn let us consider the temperature behavior of the
relation between $H_{min}$ and $H_{sat}$. According to
Eq.\ref{min} and Eq.\ref{kkamimur}
\begin{eqnarray}
\label{Hmin} H_{min} = 2g_ME_B r_{min}^{1/2}a \frac{T^2}{(g\mu_B)^2
CF}\frac{r_h^{5/2}e}{c\hbar} \propto T^{7/6}\hskip1cm ({\rm Mott
\hskip0.5cm law})
\nonumber\\
H_{min} =2\frac{\kappa^2}{e^4}r_{min}a\frac{T^2}{(g\mu_B)^2
CF}\frac{r_h^2e}{c\hbar} \propto T \hskip1cm ({\rm ES\hskip0.5cm
law})
\end{eqnarray}
At the same time $H_{sat} \propto T^{1/2}$ for Mott law and $H_s
\propto T^{3/4}$ for ES law. Thus the ratio $H_{min}/H_{sat}$
decreases with temperature decrease and this decrease is more
pronounced for Mott law.

\section{Discussion} 
\begin{figure}[htbp]
    \centering
       \includegraphics[width=0.7\textwidth]{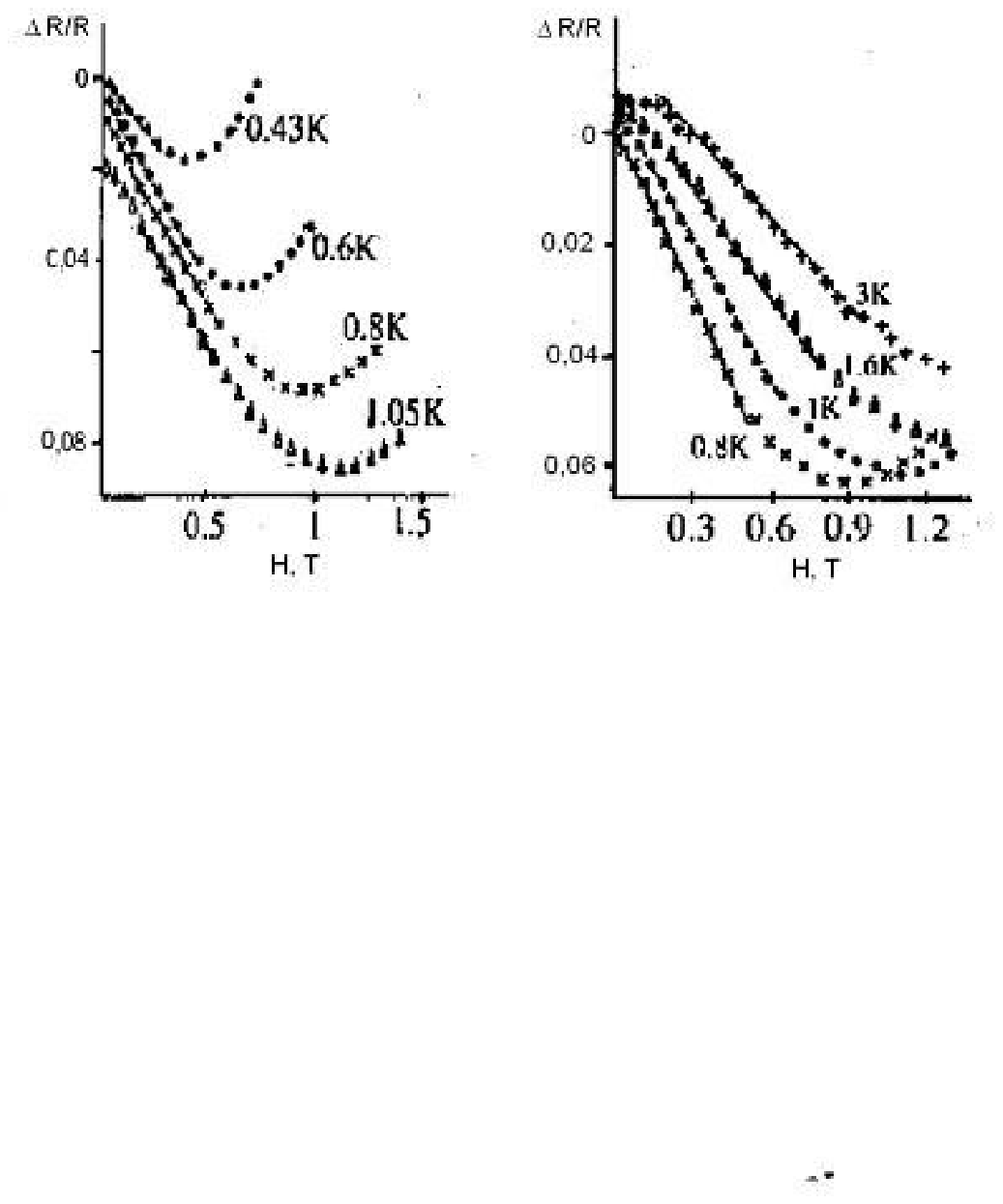}
        \caption{
        Temperature behavior of magnetoresistance for bulk
        CdTe crystals doped by donor impurities (Cl). Fig. \ref{fig1},a - the curves
        for the sample in the Coulomb gap regime, Fig.\ref{fig1}, b - for the sample
        in the Mott regime.
        \label{fig1}   }
\end{figure}

At Fig.\ref{fig1} we present our experimental
results from Ref. \cite{ours1} for 3D hopping concerning temperature
behavior of magnetoresistance for regimes of Coulomb gap hopping (
Fig. \ref{fig1},a) and of Mott-type hopping (Fig. \ref{fig1},b). It
is seen that these results are in in a qualitative agreement with
predictions of Sec.2. In particular, the minimal value of resistance
increases with temperature decrease for Mott type hopping and
decreases for the Coulomb gap hopping.  As it was noted in the
Introduction, the agreement was strongly improved when we had taken
into account more subtle spin effects \cite{ours2}, however here we
will not go into these details discussed earlier.

\begin{figure}[htbp]
    \centering
        \includegraphics[width=0.6\textwidth]{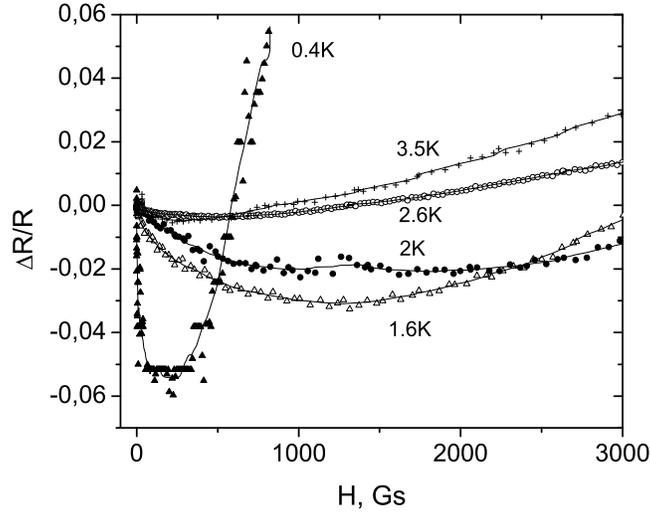}
        \caption{
        Temperature behavior of resistance for the structures of
        10 GaAs wells of thickness 15 nm, separated by AlGaAs barriers with
        thickness 15 nm. The central parts of both wells and barriers were
        doped by p-type impurity Be with concentration $1\cdot 10^{17}$
        cm$^-3$.
        \label{fig2}   }
\end{figure}

\begin{figure}[htbp]
    \centering
        \includegraphics[width=0.6\textwidth]{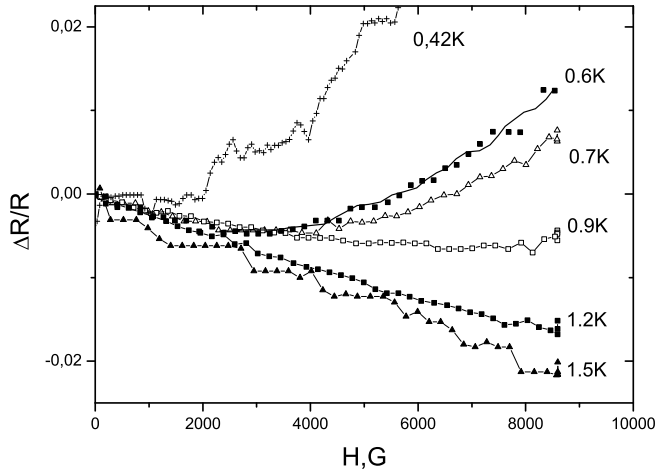}
        \caption{
        Temperature behavior of magnetoresistance  for the
        structures similar to described at Fig. \ref{fig2} but with concentration of
        Be $9 \cdot 10^{17}$ cm$^{-3}$.
        \label{fig3}   }
\end{figure}

At Figs. \ref{fig2},\ref{fig3} we present experimental results
described in \cite{ours3}, \cite{ours4}  for p-GaAs/AlGaAs:Be
multiple quantum well structures with different dopant
concentration $n$. It is seen that for the sample with smaller
concentration (Fig.\ref{fig2}) the negative magnetoresistance is
strongly enhanced with temperature decrease while for the sample
with larger dopant concentration (Fig.\ref{fig3}), in contrast, it
is suppressed with a temperature decrease.

To our opinion, the difference of magnetoresistance curves for
samples with different $n$ is related to the following fact. The
sample with smaller concentration is far from the metal-insulator
transition and the localization length is relatively small, $a
\sim 10 nm$. Thus, in a view of small $n$ and small $a$ the 3-cite
approximation for interference contribution holds, $nr_h(a
r_{min})^{1/2} \leq 1$. For heavily doped sample $a \sim 20 nm$
and $n \sim 10^{12}$cm$^{-2}$, correspondingly, $nr_h (a
r_{min})^{1/2} > 1$. As a result, the 3-cite approximation for
this sample does not hold and the interference loop includes large
number of scatterers. As it was noted above, the spin statistical
factor for each additional site with non-zero spin at $H = 0$ for
acceptor impurities is $\sim 1/4$. Correspondingly, the
interference contribution for loops involving many intermediate
scatterers vanishes at $H = 0$. As a result, the linear
contribution to negative magnetoresistance is, in any case, much
smaller than for weakly doped samples. In contrast, the quadratic
positive magneroresistance resulting from the statistical factor
given by Eq.  \ref{statquadrat} strongly increases with
temperature decrease,
\begin{equation}
\propto T^{-7/3} \hskip1cm (Mott \hskip0.2cm law), \hskip1cm
\propto T^{-9/4} \hskip1cm (ES \hskip0.2cm law)
\end{equation}

In addition, we can expect that for the sample with large $n$
$H_{sat} < H_{min}$, thus it is $H_{sat}$ which plays a role of
$H_{min}$. Due to weak temperature dependence of $R_{sat}$ the
temperature behavior at the fields larger than $H_{sat}$, that is
corresponding to the minimum of $\rho(H)$, is completely
controlled by the spin PMR which gives $\rho(H) \propto
T^{-\alpha}$ with $\alpha > 2$ . Indeed, an increase of resistance
by a factor of 4-5 is observed at the fields larger than 0.3 T for
the temperature variation from 0.9 to 0.4 K.

\section{Conclusions}
To conclude, we have reconsidered existing theory of hopping
magnetoresistance. We have shown that the random potential induced
by the background impurities can affect the asymptotics of the
localized states and, as a result, suppress to some extent the
negative magnetoresistance related to interference effects. We
have also generalized the theory for the case of acceptor states
in 2D structures including the effects of the upper Hubbard band.
The results obtained are in agreement with existing experimental
data. In particular, we explain the suppression of negative
magnetoresistance with temperature decrease observed earlier for
both 3D and 2D structures.

\end{document}